\pdfoutput=1
\documentclass[%
twocolumn,
longbibliography,
prb
]{revtex4-1}

\usepackage{graphicx}% Include figure files
\usepackage{dcolumn}% Align table columns on decimal point
\usepackage{bm}% bold math
\usepackage{xcolor} %AZ
\usepackage{braket} %AZ
\usepackage{amsmath} %AZ

\usepackage{hyperref}
\hypersetup{colorlinks,breaklinks,linkcolor=Blue,urlcolor=Blue,anchorcolor=Blue,citecolor=Blue}
\usepackage{color}
\definecolor{DarkBlue}{rgb}{0.0,0.08,0.45}
\definecolor{Blue}{rgb}{0.0,0.0,1.0}
\definecolor{Red}{rgb}{1.0,0.0,0.0}
\definecolor{RedOrange}{rgb}{0.9,0.0,0.2}
\definecolor{dgrn}{RGB}{0,150,0}
\definecolor{dgray}{gray}{0.3}
\newcommand*{\citen}{}% generate error, if `\citen` is already in use
\DeclareRobustCommand*{\citen}[1]{%
  \begingroup
    \romannumeral-`\x % remove space at the beginning of \setcitestyle
    \setcitestyle{numbers}%
    \cite{#1}%
  \endgroup
}

\begin{document}

%\preprint{APS/123-QED}

\title{\textbf{Towards \textit{ab initio} realizations of Collins' Conjecture} 
}% 

\author{Abdulrahman Y. Zamani}
\author{Kevin Carter-Fenk$^\ast$}%
%\email{kay.carter-fenk@pitt.edu}
\affiliation{%
Department of Chemistry, University of Pittsburgh,
Pittsburgh, Pennsylvania 15260, USA
}%

%  \homepage{http://www.Second.institution.edu/~Charlie.Author}
% \affiliation{
%  First affiliation for this author
% }%
% \affiliation{
%  second institution for this author
% \affiliation{%
%  Authors' institution and/or address\\
%  This line break forced with \textbackslash\textbackslash
% }%

% \collaboration{CLEO Collaboration}%\noaffiliation

\date{\today}% It is always \today, today,

% 274 words in abstract + acknowledgements
\begin{abstract}
\begin{center}
$^\ast$kay.carter-fenk@pitt.edu
\end{center}
\noindent An \textit{ab initio} approach formulated under an entropy-inspired repartitioning of the electronic Hamiltonian is presented. This \textit{ansatz} produces orbital eigenvalues each shifted by entropic contributions expressed as subsets of scaled pair correlation energy terms present in second-order M{\o}ller-Plesset (MP) perturbation theory. Under the auspices of Collins' conjecture, which suggests that the electron correlation energy is approximately proportional to the Jaynes entropy of the one-electron density matrix, we introduce a parameter that controls the accuracy of the resultant one-electron density at the MP2 level. By tuning the density in a somewhat automated way, we achieve one-electron densities on par with those from full configuration interaction for single-bond dissociation. This parameter can then be used to add a Collins'-like static correlation correction to the energy functional, capturing both dynamical and nondynamical correlation effects in many-electron systems. The performance of the proposed method and its related variants approaches the accuracy of generalized valence bond theory for estimating single bond dissociation energies (BDEs) for set of small, closed-shell molecules composed of first and second row elements.  Our results hold implications for reincorporating the missing (static) correlation energy in regularized perturbation theories that is typically discarded. Finally, we propose generic BDE parameters (accurate to within 7\% on average) that could be used for strongly-correlated systems in general.
\end{abstract}

\maketitle

%\tableofcontents

%\section{Introduction}
% {{\color{blue}{ATTN: I am tracking changes. The text in blue or commented-out will likely be excised or placed in the SI to keep this under 4500 words. Going in SI: Raw data, LiH ONs with F-D, stats/scatter plot for ionic tuned $\eta$, gap widening data w/ and w/o $\eta$. So far, approx 4400 words with tables, figures, and equations}}}

Understanding the effects that arise from the correlated motions of electrons is fundamental to the study of chemistry.\cite{Dirac1929,Wigner1934,Wigner1938,Dickens1957,Tew2007} As such, a principal endeavor of research in quantum chemistry is the development of accurate and efficient models for incorporating electron correlation effects in atoms and molecules.\cite{Lowdin1958,Urban1987,Lowdin1995} While
dynamical correlation is often handled reasonably well
using computationally-efficient density functional theory (DFT) methods, the static correlation problem is not straightforwardly addressed in a computationally
affordable way.
Thus, we seek methods with low-order polynomial scaling that can at least approximately recover static correlation, affording a compromise between accuracy and computational complexity.

The concept of entropy has been applied as an information-theoretic metric of correlation strength in many-electron systems for many years.\cite{BBM1975,Gadre1985a,Gadre1985b,Gadre1985c,Mahajeri2009,FloresGallegos2021,Salazar2023} The Shannon entropy, for example, can provide insight into the contour of electronic densities, orbital shapes, wavefunction quality, and relativistic effects. \cite{Corzo2011,Corzo2023,Lin2015,Ho1994,HoSagar1994,Ho1998a,Ho1998b,Ho1998c,Materia2024,Saha2020}
Beyond this, the entropy of the one-electron density matrix (ODM) has been suggested as a useful proxy for
static correlation in chemical systems.\cite{Martinez2023}

In a paper by Collins,\cite{Collins1993} a soft connection between $E_{\mathrm{corr}}$ and the Jaynes entropy\cite{Jaynes1957a,Jaynes1957b} $S_{\mathrm{Jaynes}}$ of an $N$-representable\cite{Coleman1963} ODM was proposed. $S_{\mathrm{Jaynes}}$ may be defined in terms of the eigenvalues of the ODM or its spin-integrated complement, the charge density matrix (CDM):
\begin{align}
S_{\mathrm{Jaynes}} = - \sum_{i} n_{i} \ln n_{i}
\label{SJaynes}
\end{align}
where eigenvalues $n_{i}$ are natural (spin-)orbital occupation numbers\cite{Lowdin1955} (NOONs) that collectively sum to the number of electrons $N$. Collins' conjecture then relates the correlation energy and Eq.~\ref{SJaynes} up to a positive multiplicative constant $\eta$:

\begin{align}
    E_{\mathrm{corr}} \approx -\eta S_{\mathrm{Jaynes}}.
    \label{Collins}
\end{align}

Given the above supposition, the correlation energy can also be seen as a measure of information or uncertainty in the ODM.\cite{Sagar2002} In the context of quantum chemistry, $S_{\mathrm{Jaynes}}$ is ideally maximized as  $E_{\mathrm{corr}}$ is minimized while approaching the true ODM. In other words, the optimal characterization of a many-particle quantum state
%is said to follow
follows the maximum entropy principle (MEP). Under a defined set of constraints or degrees of freedom,
%(nuclear and electronic coordinates, quantum numbers, basis set, etc.\ ),
it follows from the MEP that the probability distribution function (i.e.\ the density) that best describes the physical state of interest is the one gives the largest entropy or uncertainty.\cite{Jaynes1957b} 

The occupation numbers (ONs) $f_{i}$  are binary at the Hartree-Fock (HF) level, with $f_{i} = 1$ for occupied molecular orbitals (MOs) and $f_{i} = 0$  for virtual MOs. This begets a net-zero $S_{\mathrm{Jaynes}}$ and $E_{\mathrm{corr}}$ in Eqs.\ \ref{SJaynes} and \ref{Collins}, which is consistent with the notion that the HF determinant represents an
%in Roothaan HF, one determinant is sufficient information or 
%a sufficient wavefunction representation for describing an 
uncorrelated many-fermion state.
A correlated state features a non-idempotent density
matrix with
%With a non-idempotent ensemble density matrix constructed over a span of correlated electronic configurations,
fractional NOONs ({\em i.e.} $n_{i} \in [0,1]$)---yielding a non-zero $S_\mathrm{Jaynes}$.
%\footnote{$S_\mathrm{Jaynes}$ can be negative if the NOONs $n_{i}$ from a spin-traced charge density, which have a range of $[0,2]$, are used.}

Recent method developments have applied different renditions of Collins' conjecture by stratifying ONs to introduce some form of the correlation entropy term $-\eta S$ at mean-field cost. These methods largely rely on two main concepts: Fermi-Dirac (F-D) statistics\cite{Dirac1926} and density cumulants.\cite{Ziesche1995,Kutzelnigg1999} Methods based on one-electron reduced density matrix theory (1-RDMFT) %developed by Gibney et al.\
apply   corrections to self-consistent field (SCF) solutions from density cumulants.\cite{Gibney2022,Gibney2023,Gibney2024} Thermally-assisted-occupation DFT\cite{Chai2012} (TAO-DFT) and ``information-theoretic" density matrix functional
theory\cite{Wang2022,Liu2024} (i-DMFT) utilize the F-D function to obtain ONs: 
\begin{align}
    f_{i} = (1+e^{({\varepsilon_{i} - \mu})/{\eta}})^{-1}
\end{align}
where $0 \leq f_{i}\leq 1$, $\varepsilon_{i}$ is the eigenvalue for orbital $i$, and $\mu$ is a chemical potential chosen to conserve $N$ electrons. Here, the ``correlation temperature" $T$ is taken to be $\eta$ in energy units or Hartrees (Ha).
%and Boltzmann constants $k_{B}$ are considered apart of the atomic unit system.
The F-D smearing function is symmetrical and does not directly consider the features of the density matrix, rather, the chemical potential dictates the extent of ON stratification across occupied and virtual single-particle states.
%This type of Fermi ``smearing" is done symmetrically about $\mu$ and can lead to highly degenerate orbital occupations with increasing $\eta$.
Furthermore, though chosen for its physical relevance, the F-D distribution is not unique as Gaussian or linear distributions can also be used.
%perform smearing in a similar way while maintaining that $0 \leq f_{i}\leq 1$.
Nonetheless, these ONs are then used to compute the electronic entropy deduced from statistical thermodynamics
\begin{align}
    S_\text{F-D} = - \sum_{i} f_{i} \ln f_{i} + (1-f_{i}) \ln (1-f_{i})
    \label{fdEntropy}
\end{align}
rather than $S_{\text{Jaynes}}$, though $n_{i}$ could also be used in the above expression.\cite{FloresGallegos2023} Note that the form of $S_\text{F-D}$ resembles that of a metric known as the particle-hole symmetric correlation entropy.\cite{GoriGiorgi2002,Gottlieb2005,Gottlieb2007}
%or ``nonfreeness" of an uncorrelated electronic subsystem that is superadditive and shares the same density operator as the reference system---without any specific reference to $E_\mathrm{corr}$.

Invoking $T$ and $\mu$, both statistical mechanical concepts, is an interesting choice for atoms or molecules typically modeled at 0 Kelvin, as they are not well-defined for such finite systems.\cite{Baerends2022} Nevertheless, the objective within thermal SCF\cite{Mermin1963,Sokoloff1967,Kaplan1975,Trickey2012} is to minimize the electronic Helmholtz free energy 
\begin{align}
    A \equiv E_\mathrm{SCF} - TS_\text{F-D}
    \label{freeEnergy}
\end{align}
which is an analog form for expressing $E_0$ with some contribution from $E_\mathrm{corr}$
\begin{align}
    E_0 \approx E_\mathrm{HF} - \eta S_\text{F-D}.
    \label{gcEnergy}
\end{align}
The essence of i-DMFT and TAO-DFT is captured in Eq.\ \ref{gcEnergy} for a given $\eta$.
%In reformulated versions of the latter method (rTAO), the ``zero-temperature'' canonical HF or DFT solution is converged and the iteratively determined $- \eta S_\text{F-D}$ term is a {\em post hoc} correction to the energy.
A caveat in i-DMFT pertaining to the correlation or cumulant energy is 
\begin{align}
    E_{\mathrm{corr}} = -\eta S_{\text{F-D}}-b
 \label{idmftEnergy}   
\end{align}
where $b$ is an empirical intercept that keeps $E_{\mathrm{corr}}$ somewhat linear with the choice of entropy; in this case $S_{\text{F-D}}$ is used.\cite{Irimia2023} 
One challenge these methods face is the determination\cite{Lin2017,Chen2022} of $\eta$, since it is not universal in magnitude and is often fit (along with $b$ in Eq.\ \ref{idmftEnergy}) to high-level correlated wavefunction methods. The ONs obtained through the F-D function with a given $\eta$ can become fractional rapidly which is very useful for getting proper ONs at the homolytic molecular dissociation limit. However, this approach is also unstable for asymmetric molecules such as LiH given a correlation temperature that would in turn supply meaningful energy corrections (see Supplementary Material). This is not surprising, as the F-D distribution is symmetric and the function that transforms the canonical MOs to the basis of natural orbitals (NOs) which diagonalize the correlated ODM does not necessarily behave symmetrically.

Notwithstanding the reasonable assertions against the use of the Collins conjecture,\cite{Cioslowski2024} the usefulness of information-theoretic correlation metrics in chemistry and physics should not be overlooked,\cite{Guevara2005,AlivertiPiuri2024,Esquivel2015,Huang2005,Ding2020,Martinez2023,Isobe2002,Amovilli2004,Sagar2008,Feng2025}
nor should the corpus of previous numerical studies supporting Eq.~\ref{Collins}.\cite{Esquivel1996,Ramirez1997,Ramirez1998,Ziesche1999,Sagar2002}
Motivated by the prospect of including static correlation within single-reference approximations, but looking to move beyond {\em ad hoc} approximations such as F-D smearing in molecular systems, we provide a more solid foundation for such entropy-inclusive approaches that is capable of describing asymmetrical ODMs ({\em e.g.} ODMs of LiH dissociation or at point-defects in solids). Herein, we present an {\em ab initio} approach for realizing Collins' conjecture in terms of renormalized wave function amplitudes and test our approach on the strong-correlation problem of bond breaking.
%Motivated by the overall prospects of entropic renormalizaton of single-reference methods, we present an \textit{ab initio} approach for realizing Collins' conjecture and assess its performance on the strong correlation problem of bond-breaking. 

%\section{Theory}

We begin with the M{\o}ller-Plesset (MP) partitioning of the electronic Hamiltonian $\mathcal{H}$:
\begin{align}
\mathcal{H} = \mathcal{H}_0 - \lambda \mathcal{W}
\label{partMP}
\end{align}
where $\mathcal{H}_0$ is the unperturbed Hamiltonian taken to be the sum of Fock operators, $\mathcal{W}$ is the fluctuation potential, and $\lambda$ is the usual dimensionless coupling-strength parameter. The lowest-order correlation correction to the HF energy occurs through matrix elements between the zeroth-order wavefunction and doubles substitutions generated with the cluster operator:
\begin{align}
    \mathcal{T}_2 = \frac{1}{4}\sum_{ijab} t_{ij}^{ab} a_{a}^{\dagger} {a}_{b}^{\dagger} {a}_{j} {a}_{i}
    \label{t2}
\end{align}
where the set of $\{i,j,k,\ldots\}$ and $\{a,b,c,\ldots\}$ refer to occupied and virtual MO indices respectively. The corresponding energy and amplitude expressions are, 
\begin{align}
    E^{(2)}_{\mathrm{MP}} = \frac{1}{4}v_{ab}^{ij}t_{ij}^{ab}
    \label{E2}
\end{align}
and
\begin{align}
        \Big[\big(f_c^a\delta_d^b+\delta_c^af_d^b\big)\delta_i^k\delta_j^l-\big(f_i^k\delta_j^l+\delta_i^kf_j^l\big)\delta_c^a\delta_d^b\Big]t_{kl}^{cd} = -v_{ij}^{ab}
\end{align}
%
%\begin{align}
%    t_{ij}^{ab} = \frac{\braket{ab||{ij}}}{\varepsilon_{i} + \varepsilon_{j} - \varepsilon_{a} - \varepsilon_{b}}
%    \label{tijab} 
%\end{align}
%
where $v_{ij}^{ab}=\langle ij||ab\rangle$ are anti-symmetrized electron repulsion integrals, and $f_p^q$ are one-electron Fock matrix elements.
By virtue of Collins' conjecture, we enforce the following equality, taking $E^{(2)}_{\mathrm{MP}}  \approx  E_\mathrm{corr}$:
\begin{align}
 E^{(2)}_{\mathrm{MP}} = - \eta S .
 \label{etaMP2collins}
\end{align}
The total entropy is separated in terms of one-electron entropies over general $\{p,q,r,\ldots\}$ orbital indices,
\begin{align}
    S = \sum_p s_p  = \sum_i s_i + \sum_a s_a
\end{align}
which can then be inserted into Eq.~\ref{etaMP2collins}, 
\begin{align}
     \eta \left(\sum_i s_i + \sum_a s_a\right) = \frac{1}{4} v_{ab}^{ij}t_{ij}^{ab}
\end{align}
This form suggests a relationship between orbital-by-orbital one-particle corrections to the energy that manifest due to electronic entropy and the correlation energy. These one-particle contributions may be written directly as tensor contractions over the $t$-amplitudes
\begin{align}\label{etaMP2expand}
     \eta \bigg( \sum_i s_i + \sum_a s_a \bigg) = \frac{1}{8}\bigg(\sum_{i} \underbrace{v_{ab}^{kj}t_{ik}^{ab}\delta_{ij}}_{\eta s_i} \ - \sum_{a} \underbrace{v_{cb}^{ij}t_{ij}^{ac}\delta_{ab}}_{\eta s_a}
      \bigg).
\end{align}
Specifically, this form motivates the use of scaled one-particle correlation energies
to account for (or tune) entropy in correlated wave function calculations.
%We see that Eq.\ \ref{etaMP2collins} is satisfied up to a multiplicative constant in Eq.\ \ref{etaMP2expand}.
%Note that the terms on the RHS of the above equation can be viewed as correlation corrections to the ionization potentials (IP) and electron affinities (EA): 
%
%\begin{align}
%   \frac{1}{4}\bigg(\sum_{i} \underbrace{\bigg(\frac{1}{2}{\sum_{kab}  \braket{ik||{ab}}t_{ik}^{ab}}\bigg)}_{\mathrm{IP}^\mathrm{corr}_i} \ +   \sum_{a} \underbrace{\bigg(\frac{1}{2} {\sum_{ijc}  \braket{ij||{ac}}t_{ij}^{ac}}\bigg)}_{\mathrm{EA}^\mathrm{corr}_a}
%      \bigg). 
%     \label{etaMP2IPEA}
%\end{align}
%
%The correlation terms for ${\mathrm{IP}^\mathrm{corr}}$ and ${\mathrm{EA}^\mathrm{corr}}$ may be interpreted as pair removal energies and pair relaxation energies respectively.\cite{Pickup1973,Szabo} 

 %The relationship between the individual orbital entropies presumes nothing about the form of the entropy functional that correlates the orbitals.
Collins' suggestion implies that $-\eta S_\mathrm{Jaynes}$ engenders the true many-body entropy and the entirety of $E_\mathrm{corr}$. However, when the ONs used resemble the NOONs of a ODM that corresponds to the actual wavefunction, the total one-electron entropy should be adept at describing nondynamical (or static) correlation.\cite{Sinanoglu1963, ShaJiaPav23} The remaining instantaneous electrostatic repulsions encompass dynamical correlation effects and are aptly modeled with
%pair correlation terms from
MP2 as a first-approximation.\cite{Sinanoglu1969} 

Modifying $\mathcal{H}_0$ in the partitioning of Eq.\ \ref{partMP} to include correlation corrections to the orbital eigenvalues gives:
\begin{align}\label{etaH0}
    \mathcal{H}_0 &= \sum_p \bigg(\mathcal{F}_{pp}  a_{p}^{\dagger} a_{p} - \eta  s_p\bigg) \nonumber \\
    &\approx \sum_{ij}\bigg(f_i^j+\eta v_{ab}^{kj}t_{ik}^{ab}\bigg)a_j^\dagger a_i+\sum_{ab}\bigg(f_b^a-\eta v_{cb}^{ij}t_{ij}^{ac}\bigg)a_b^\dagger a_a \nonumber \\
    &= \sum_{ij}F_i^ja_j^\dagger a_i+\sum_{ab} F_b^a a_b^\dagger a_a
\end{align}
leading to the modified amplitude expression,
\begin{align}
    \Big[\big(F_c^a\delta_d^b+\delta_c^aF_d^b\big)\delta_i^k\delta_j^l-\big(F_i^k\delta_j^l+\delta_i^kF_j^l\big)\delta_c^a\delta_d^b\Big]t_{kl}^{cd} = -v_{ij}^{ab}
\end{align}
In the appropriate basis, Eq.~\ref{etaH0} is diagonal and the amplitude expression becomes
isomorphic with MP2, albeit with a set of dressed
orbital energies that contain correlation.\cite{SheHenScu14}
Note that the resultant amplitude equations are equivalent to a modification of
the coupled cluster doubles (CCD) equations such that they only include (scaled) mosaic terms
(the threefold contractions in Eq.~\ref{etaH0}). Finally, the resultant $\eta$MP2 approach becomes equivalent to mosaic CCD when $\eta=1/2$. We will later find that the ideal
$\eta=0.4$~Ha does not stray far from this value.
%
%\begin{align}
%    \mathcal{H}_0 &= \sum_p \bigg(\mathcal{F}_{pp}  a_{p}^{\dagger} a_{p} - \eta  s_p\bigg) \nonumber  \\ &\approx \sum_{i} \bigg( \mathcal{F}_{ii} a_{i}^{\dagger} a_{i} +\eta {\sum_{kab}  \braket{ik||{ab}}t_{ik}^{ab}a_{i}^{\dagger} a_{i}} \bigg) \nonumber \\ &+ \sum_{a} \bigg( \mathcal{F}_{aa} a_{a}^{\dagger} a_{a} - \eta {\sum_{ijc}  \braket{ij||{ac}}t_{ij}^{ac}a_{a}^{\dagger} a_{a}}
%      \bigg). 
%      \label{etaH0}
%\end{align}
%
%The $\mathcal{F}_{pp}$ are the orbital energies $\varepsilon_{p}$. As the entropy-corrected $\varepsilon_{p}$ depend on the amplitudes attenuated by $\eta$, the resultant equations must be solved iteratively and may use canonical MP2 amplitudes an initial guess.

%However, only the diagonal elements of $\mathcal{H}_0$ are influenced by this shift, %So,
%unlike size-consistent and orbital invariant second-order Brillouin-Wigner methods\cite{CarterFenk2023a,CarterFenk2023b,Dittmer2025} (BW-s2) to which this entropy-driven correction bears resemblance.

The presented $\eta$MP2 approach is orbital-invariant (oi), size-consistent, and size-extensive with computational complexity of iterative $\mathcal{O}(N^{5})$ as the amplitude equations are self-consistent.\cite{Szabo1977}
%The scaling of the  oi-$\eta$MP2 calculation is formally $M\times\mathcal{O}(N^{5})$, where $M$ is
%the number of iterations required to converge the
%self-consistent amplitude equations. 
We must now determine the parameter $\eta$, which we take to control the quality of the predicted ODM and associated ONs (thus tuning the entropy). Alongside $\eta$, these ONs may be used to add the static correlation correction provided by $-\eta S_{\text{Jaynes}}$ to the total free energy. As $\eta$ modulates our proxy for static correlation, it is sensible to determine $\eta$ in cases of homolytic bond cleavage of singlet molecules (e.g.\ stretched $\mathrm{H}_2$), where the eigenvalues of the highest occupied (HOMO) and lowest unoccupied (LUMO) MOs become degenerate and static correlation tends to dominate.

%This method is denoted as $\eta \mathrm{MP2}$.
%The form of $\mathcal{H}_0$ in Eq.\ \ref{etaH0} is not orbital invariant but can be made so through careful accounting of the tensorial components of each term.\cite{White1998} In principle, $\eta \mathrm{MP2}$ is isomorphic to canonical MP2, permitting the reinheritance of the same properties such as orbital invariance,  size-consistency, and size-extensivity\cite{Szabo1977}---all while embracing the connection between the information entropy, corrections to the HF density, and thus the electronic energy. For simplicity, we maintain the choice of Eq.\ \ref{etaH0} as a proof of concept. What is left to determine is the parameter $\eta$. In homolytic bond cleavage of a singlet molecule (e.g.\ stretched $\mathrm{H}_2$), the eigenvalues of the highest occupied (HOMO) and lowest unoccupied (LUMO) MOs become degenerate. Not only is single-reference restricted HF incapable of dissociating the system into a diradical, but MP2 will also diverge due to small denominators caused by these orbital degeneracies. Here, static correlation is imbued in the $\eta \mathrm{MP2}$ \textit{ansatz} via the entropy, leaving $\eta$ as a parameter to be determined.
%and $\eta$ must be chosen to widen the HOMO-LUMO gap at the dissociation limit. 
 
A reasonable choice of $\eta$ is one that recovers the correct ONs of the frontier orbitals participating in the molecular bond. For example, in single-bond breaking, the NOONs corresponding to the HOMO and LUMO should equate closely to 1. For a stretched system, the $\eta \mathrm{MP2}$ ODM that gives qualitatively correct NOONs should be superior to the restricted (R)MP2 ODM which can yield negative occupations, violating the positive-definite character of an $N$-representable density matrix.\cite{Ho1998a}  In principle, selecting a value of $\eta$ that guides the density towards the correct electronic structure can be automated whilst retaining a modest fifth-order scaling bottleneck step macro-iterated to the desired precision in $\eta$. This is conceptually satisfying as it categorizes $\eta \mathrm{MP2}$ closer to what is expected of an \textit{ab initio} method capable of properly tuning the density corrections afforded by a basic pair correlation theory. This feature is juxtaposed to methods such as i-DMFT or TAO-DFT that rely on fitting to higher-level theories or quantities extraneous to finite molecular quantum systems such as chemical potential or temperature for mimicking a correlated ODM.

An especially interesting attribute
of our of $\eta$ parameterization is that
finding $\eta$ in the limit where the most important contribution to the correlation energy is
static ensures by Eq.~\ref{Collins} that $-\eta$S essentially
accounts for missing static correlation components
that are normalized away from the $\eta$MP2 energy.
The Collins' parameterization allows us to add some semblance of the ill-behaved correlation back into our renormalized MP2
to obtain $\eta$MP2$-\eta S$. This is fundamentally different from typical regularized MP2 methods that simply
eliminate problematic correlation contributions without any attempt to account for what was discarded.\cite{StuHea13,LeeHea18,SheLoiRet21,KelTsaReu22,CovTew23}
While the extent of regularization %(and thus the amount of dynamical correlation)
is tuned to obtain the correct electronic structure of the system, the entropy
reintroduces static correlation that is usually eliminated by other approaches.

Finally, we note that we merely use Collins' conjecture as a guidepost and
there is no mathematical reason to impose the burden of tuning the ODM and
modulating the static correlation energy onto a single parameter, $\eta$.
Much like RDMFT, which uses separate parameters for accomplishing F-D smearing
and scaling the resultant entropy, we will also consider a second parameter $\gamma=\alpha\eta$,
where $\alpha$ is a linear scaling parameter that we determine by best-fit to the
bond dissociation and total energies for stretched bonds as described below.\cite{Gibney2022,Gibney2023,Gibney2024}
This results in a total energy $E=E_{\eta\text{MP2}}-\gamma S_{\text{Jaynes}}$,
where $\eta$ is tuned to optimize the ODM and resultant entropy
and $\gamma$ modulates the static correlation energy. %In practice,  $\gamma$ suppresses over-correlation from the entropy term. 

%\section{Results \& Discussion}

To test the performance of $\eta \mathrm{MP2}$, bond dissociation energies (BDEs or $D_{e}$) are computed from total energies. An assortment\cite{Kraka1992a,Kraka1992b,Feng2003} of closed-shell molecules containing H and $p$-block elements are selected. Whereas double or triple-bond breaking requires quadruple or sextuple excitations,\cite{Krylov1998,Musial2005} MP2 contains only double substitutions.
Consequently, only single-bonded systems are considered in this study.
%Only BDEs for single bonds are chosen since the excitation space of MP2 spans only disconnected doubles.
%Double or triple bond breaking would require access to quadruple or sextuple excitations present in higher-order methods \cite{Krylov1998,Musial2005} and are not included this study.
The results for $\eta \mathrm{MP2}$ along with related $\eta$-based methods: $\eta \mathrm{MP2}{-}\eta S $ and $\eta \mathrm{MP2}{-}\gamma S $, where $S$ is shorthand for the Jaynes entropy, are compared to ``experimental", CCSD, and CCSD(T) values.
%Specific computational details on basis sets, geometries, and how $D_{e}$ are inferred from reference data pertaining to $D_0$ are available in the Supplementary Material.
All $\eta \mathrm{MP2}$ calculations were performed with a development version of Q-Chem~v6.2.\cite{qchem} 
% %
% \begin{figure}
%     \centering
% \includegraphics[width=\linewidth]{etaMP2diradical-oi-actual.png}
%     \caption{Average errors in kcal/mol of  orbital-invariant (oi) $\eta$MP2 with respect to total energies, theoretical BDEs, and experimental BDEs. The parameter $\eta$ is tuned to provide  HOMO-LUMO occupations close to 1 for fragments stretched to 100\r{A}. {\color{blue}{Here, the value of $\eta$ in $\eta S$ is made proportionate, by a factor of $1/8$, to $\eta$ in $\eta S$ determined without orbital invariance.}} }
%     \label{fig:oi-eta-diradical}
% \end{figure}
%

\begin{figure}
    \centering
\includegraphics[width=\linewidth]{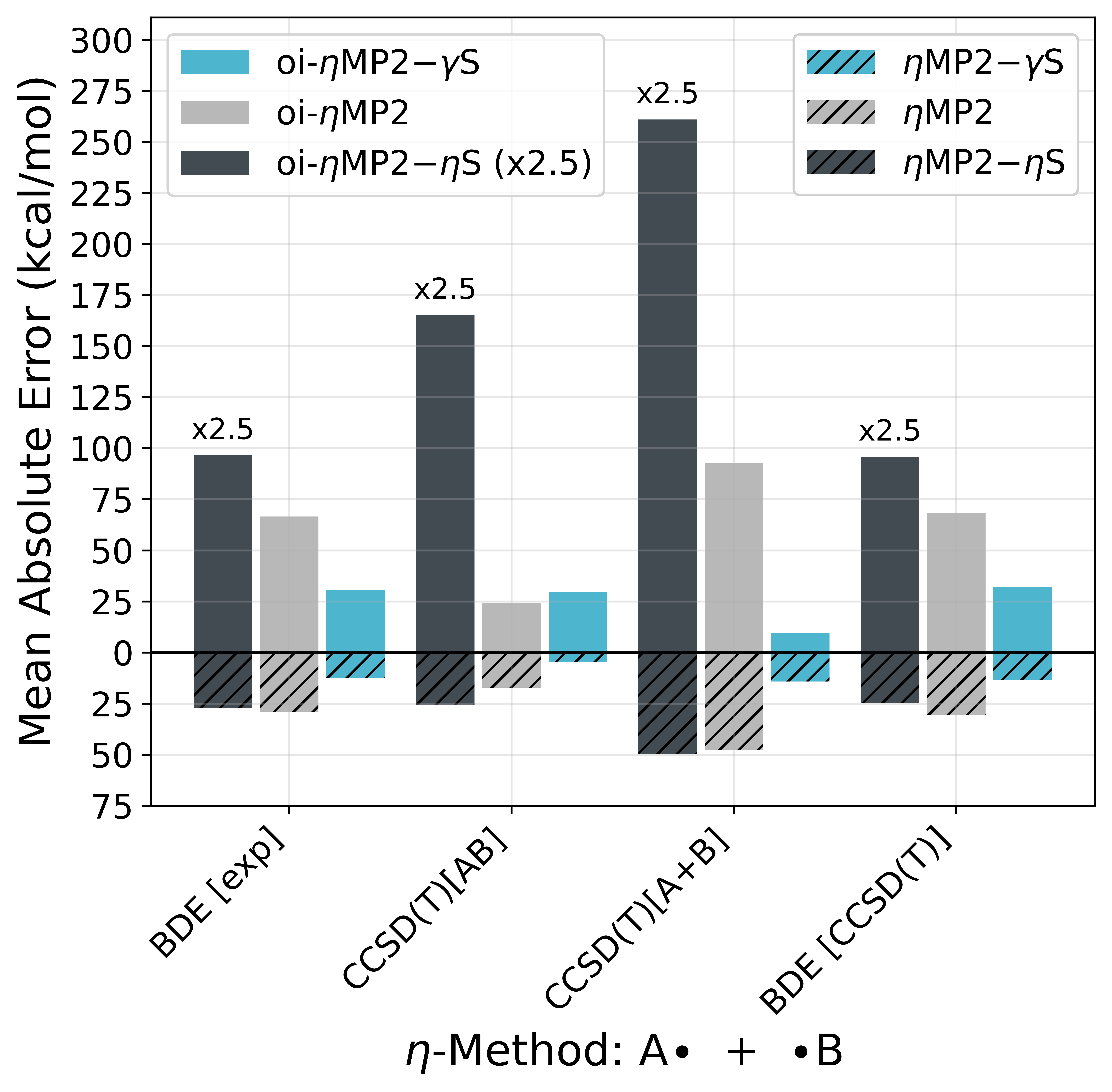}
    \caption{Average errors in kcal/mol of  orbital-invariant (oi) and non-invariant $\eta$MP2 with respect to CCSD(T) total energies, CCSD(T) BDEs, and experimental BDEs. The parameter $\eta$ is tuned to provide  HOMO-LUMO occupations close to 1 for fragments stretched to 100\r{A}. The optimal value of oi and non-oi $\gamma$ is $\eta$ scaled by 0.123 and 0.500 respectively.}
    \label{fig:oi-eta-diradical}
\end{figure}
%
%
%
% {\color{blue}{The input parameter $\eta$, with a mantissa of up to $10^{-4}$, is chosen such that the difference between the HOMO and LUMO $\eta \mathrm{MP2}$ NOONs at 100\r{A} is no larger than 0.02. The optimal $\eta$ for a fragmented molecule is then used on the same molecule at equilibrium. The value of $S_\mathrm{Jaynes}$ is computed with sum of up-spin $\alpha$ and down-spin $\beta$ $s_p$ values and applied to $\mathrm{HF}{-}\eta S$ and $\eta \mathrm{MP2}{-}\eta S $ along with the same individually optimized entries for $\eta$. We note that the magnitudes of $\eta$ are similar to that of the analogous $\theta$ or $\kappa$ parameter in aforementioned TAO and density matrix methods.}}

Results displayed in Fig.\ \ref{fig:oi-eta-diradical} suggest that the inclusion of  $-\gamma S$ corrections with (oi) $\eta \mathrm{MP2}$ exhibit better performance at the dissociation limit than at equilibrium, where the $\eta \mathrm{MP2}$ and  $\eta \mathrm{MP2}-\gamma S$ methods perform similarly. With respect to experimental and CCSD(T) BDEs, the mean absolute errors (MAEs) are the lowest with the addition of $-\gamma S$ ($\sim$30 kcal/mol). The use of $\eta \mathrm{MP2}$ without entropy corrections gives BDE errors about twice as a large. Incorporating unscaled $-\eta S$ corrections results in wholly unsatisfactory errors. The average value of $\eta$, tuned to the correct NOONs, is about 384 mHa, which is similar to the {\em ab initio} 500~mHa in mosaic CCD. However, the average value of $\gamma$, which controls the magnitude of the entropy-based static correlation is 45 mHa, which falls within the typical range of the analagous parameters used in TAO and iDMFT (10-100 mHa).

These results do not reach chemical accuracy, but the  $\eta \mathrm{MP2}{-}\gamma S$ BDEs are of similar quality to generalized valence bond (GVB) theory.\cite{Kraka1992a}  We note that equilibrium  $\eta \mathrm{MP2}-\gamma S$ energies are similar to canonical MP2, even with the value of $\eta$ that was selected for dissociation, as the role of static correlation is much smaller at equilibrium.
The role of the $\eta$ parameter is most significant at the dissociation limit as it quells divergences typical of RMP2 by widening the HOMO-LUMO gap (see Supplementary Material for details). This temperament of the ODM leads to suppressing of dynamical correlation while the ODM-derived entropy (and thus static correlation) is enhanced at large $R$.

% The total energies at long distance computed with $\eta \mathrm{MP2}$ undershoot the CC energies by up to $\sim$200 mHA while  $\eta \mathrm{MP2}{-}\eta S$ overshoot the CC energies by up to $\sim$60 mHa. For equilibrium total energies, $\eta \mathrm{MP2}$ undershoots up to $\sim$70 mHa and $\eta \mathrm{MP2}{-}\eta S$ overshoots up to $\sim$120 mHa. While $\mathrm{HF}{-}\eta S$ improves upon RHF, the energy estimates, given a fixed $\eta$, are unsatisfactory. 

As we forgo invariance once we add the entropy correction regardless (the NOONs and thus $E_{\text{static}}[S_{\text{Jaynes}}]$ are valid only in the NO basis), we also obtained results, displayed in Fig.\ \ref{fig:oi-eta-diradical}, for non-invariant $\eta$MP2,  a simplified approximation to oi-$\eta$MP2 discussed in the Supplementary Material. Here, the average value of $\eta$, tuned to the correct NOONs, is about 65 mHa, which is already within the typical range for an entropy-scaling parameter. This suggests that this non-invariant $\eta \text{MP2}$ may be adequately corrected for static correlation by using the single $\eta$ parameter to tune the density and modulate the entropy contribution.
%This is similar to analogous parameters used in TAO and iDMFT, which lie roughly between 10 and 100 mHa. 

% The MAEs of non-variant $\eta \mathrm{MP2}$ are about  29,  33, and 31 kcal/mol when respectively compared to BDEs from experiment, CCSD, and CCSD(T). In the same order, the MAEs are about 27,  23, and 25 kcal/mol for $\eta \mathrm{MP2}{-}\eta S$. These $\eta \mathrm{MP2}{-}\eta S$ results are similar to those of the orbital-invariant version. 

With respect to CC and experiment, non-invariant $\eta$MP2 and $\eta$MP2-$\eta S$ can offer a $\sim$50\% error reduction for BDEs ($\sim$13-16 kcal/mol) compared to its orbital-invariant counterpart. Despite this apparently useful feature, the non-invariant $\eta\text{MP2}$ destroys information and cannot reliably construct potential energy surfaces which is potentially catastrophic for geometry optimizations, dynamics, vibrational frequencies, and thermochemistry.
The optimal $\eta$ parameter in non-invariant $\eta\text{MP2}$ also varies by 28\% from the mean value across our data set, precluding
a generic suggestion of $\eta$ for any BDE problem.
On the other hand, oi-$\eta\text{MP2}-\gamma S$ performs well for bond breaking, yields smooth potential energy surfaces (PESs), and the optimal
$\eta$ (thus $\gamma$ as well) varies by only 7\% from the mean value, allowing a somewhat general $\eta=384$~mHa and $\gamma=45$~mHa to be suggested.
We posit that magnitude and specificity of $\eta$ will depend on both the class of strong correlation and the many-body method of choice. This is demonstrated in Table~S3 of the Supplementary Material, which shows that $\eta\text{MP2}-\gamma S$ electric dipole moments computed at equilibrium geometries are on par with CCSD, but require a much larger $\eta$ parameter than reported here for BDEs. However, within a given class of correlation a single $\eta$ parameter is sufficient.
%While we expect $\eta$ and $\gamma$ to vary from property to property, such a tight error distribution for single-bond breaking suggests that these parameters could be used
%with oi-$\eta\text{MP2}-\gamma S$ to model the static correlation
%in systems other than bond-breaking cases or where the parameterization of $\eta$ is not obvious ({\em i.e.} polyatomic molecules that are strongly-correlated at their equilibrium geometry).
For general use, oi-$\eta \mathrm{MP2}$ along with ${-}\gamma S$ corrections for reasonable energy estimates is suggested.

\begin{figure}
    \centering
    \includegraphics{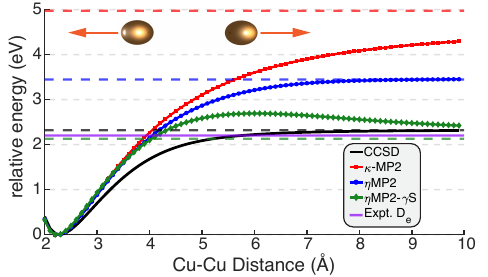}
    \caption{Potential energy curves for Cu$_2$ 
    computed using a variety of (oi) methods with the cc-pVTZ basis.
    Dissociation limits (estimated at $R=100$~\AA)
    are shown for each method as a like-colored
    horizontal dashed line. The experimental
    $D_e$ from Ref.~\citen{HerHub79} is shown as a horizontal purple line.
    }\label{fig:Cu2PES}
\end{figure}

To emphasize the generality of our chosen $\eta$ and $\gamma$ parameters for BDEs,
we examine the PES
of Cu$_2$ dissociation in Fig.~\ref{fig:Cu2PES}. The CCSD curve is taken as a reference for the potential energy surface shape, while both CCSD and experimental $D_e$ values provide references for well depth.
Without tuning the parameters for Cu$_2$, we find that HOMO and LUMO occupations are nearly equivalent at 0.96~electrons at $R=100$~\AA, suggesting that our parameter set is reasonably robust for BDEs beyond the p-block elements.
The $\eta\text{MP2}-\gamma S$ results are in best agreement with CCSD relative to PES shape and $D_e$.
The difference in $D_e$ between $\eta\text{MP2}-\gamma S$ and CCSD is 4.3~kcal\slash mol, with the comparison to experiment yielding an even smaller error of just 1.6~kcal\slash mol.
While energy differences are of primary concern, we note that at equilibrium $\eta\text{MP2}-\gamma S$ overestimates the total energy.
%We note that at equilibrium the $\eta\text{MP2}-\gamma S$ approach and other MP2-based approaches overestimate the total energy, but the energy differences are what concern us here.

Notably, $\eta\text{MP2}-\gamma S$ outperforms both $\kappa$-MP2 and $\eta$MP2 approaches for $D_e$. This is because regularization or renormalization methods temper overly large $t$ amplitudes, while $\eta\text{MP2}-\gamma S$ empirically accounts for static correlation effects at large $R$.
This is clearly reflected in
the much larger $D_e$ errors relative to experiment of 64~kcal\slash mol
and 29~kcal\slash mol for $\kappa$-MP2 and $\eta$MP2, respectively.
We have also tested the same fixed-optimal parameters on singlet iron carbonyl BDEs (see Supplementary Material). Though the character of these species at dissociation is different than that of the test set molecules due to strong $\pi$-backbonding that results in dative metal-ligand bonds,
the relative signed errors with respect to CCSD BDEs remain within the expected range at $\sim$3--29 kcal/mol.

% There is an apparent lack of error cancellation due to more prominent $\eta$-driven energy changes occurring in the fragments at 100\r{A} separation. Any over-counting of correlation is a result of the {\em ad hoc} addition of ${-}\eta S$ to $\eta \mathrm{MP2}$. 
% Again, this yields a kind of electronic free energy, which is more or less a finite temperature property.  
% Without the ${-}\eta S$ correction, the total $\eta \mathrm{MP2}$ energy at equilibrium is already quite decent compared to CCSD.

% {\color{blue}{We emphasize that $\eta \mathrm{MP2}$ energies at dissociation are computed differently than CCSD[A+B] or CCSD(T)[A+B], where the latter two methods use the sum of independent open-shell fragments. Attempting to compute fragment energies with $\eta \mathrm{MP2}$ would 1) nullify the applicability of entropic corrections for addressing strong correlation along the PES and 2) would offer ambiguous, possibly spin-contaminated perturbative density cumulant that does not lead to a proper ensemble density with definite spin. }}
%
\begin{table}[ht]
\centering
\caption{$S_\mathrm{Jaynes}$ and NOONs from the orbital-invariant (oi) formulation of RI-$\eta \mathrm{MP2}$/cc-pVTZ and CASCI/cc-pVTZ with growing active spaces $(N,M)$ for LiH calculated at equilibrium ($\mathrm{r}_\mathrm{eq}$)  and stretched ($\mathrm{r}_{100\text{\r{A}}}$) bond lengths. CASSCF/cc-pVTZ results are in \textit{italics}.}
\setlength{\tabcolsep}{5pt}  %5pt
\renewcommand{\arraystretch}{1.5} %1.50
\begin{tabular}{l|ccccccc}
\hline\hline
$\mathrm{r}_\mathrm{eq}$  & oi-$\eta \mathrm{MP2}$ & (4,4) & (4,8) & (4,16) & (4,32) & (4,44) \\ 
$S_\mathrm{Jaynes}$ & 0.382 & 0.002 & 0.025 & 0.155 & 0.425 & 0.467  \\
&  & \textit{0.252} & \textit{0.337} & \textit{0.358} & \textit{0.357} & \textit{0.354}  \\
 $n_\mathrm{HOMO}$ & 1.945 &  2.000 & 1.997 &  1.980 & 1.942 & 1.941 \\ 
 &  & \textit{1.952} & \textit{1.941} & \textit{1.941} & \textit{1.942} & \textit{1.942}  \\
 $n_\mathrm{LUMO}$ & 0.023
 &  0.000 & 0.000 &  0.003 & 0.005 & 0.004 \\
 &  & \textit{0.037} & \textit{0.031} &  \textit{0.030} & \textit{0.030} & \textit{0.029} 
 \\ \cline{1-7} 
 $\mathrm{r}_{100\text{\r{A}}}$  & oi-$\eta \mathrm{MP2}$ & (4,4) & (4,8) & (4,16) & (4,32) & (4,44) \\ 
 $S_\mathrm{Jaynes}$ & 1.660 &  1.386 & 1.386 & 1.389 & 1.631 & 1.673 \\
 &  & \textit{1.409} & \textit{1.409} & \textit{1.409} & \textit{1.409} & \textit{1.409}  \\
 $n_\mathrm{HOMO}$ & 0.982 &  1.000 &  1.000 & 1.000 & 0.935 &  0.930\\ 
 &  & \textit{1.000} &  \textit{1.000} &  \textit{1.000} &  \textit{1.000} &  \textit{1.000}  \\
 $n_\mathrm{LUMO}$ & 0.982 & 1.000 & 1.000 &  1.000 &  1.000 &  1.000 \\ 
  &  & \textit{1.000} &  \textit{1.000} &  \textit{1.000} &  \textit{1.000} &  \textit{1.000} \\
\hline\hline
\end{tabular}
\label{CAS-eta}
\end{table}

% To bring the total energy down towards CCSD or CCSD(T) effectively required  setting $\eta$ such that the HOMO and LUMO NOONs was totaled to 0 and 2 respectively, resulting in an erroneous fragment separation into ionic species. The MAEs against experimental and theoretical BDEs were lowered by about 10 kcal/mol with this approach, but is deemed inappropriate despite improvements upon the errors. We also note that the entropies calculated with NOONs from this ``ionic" picture of dissociation did not change meaningfully between equilibrium and stretched states. This is unintuitive and completely different than when $\eta$ is tuned to get the qualitatively correct $\eta \mathrm{MP2}$ density, where the $S_\mathrm{Jaynes}$ obtained from more proper NOONs is double or greater in magnitude at dissociation compared to the $S_\mathrm{Jaynes}$ at equilibrium. 

Returning to our original data set, we note that the values of $S_\mathrm{Jaynes}$ obtained with $\eta$MP2 are nearly twice or greater in magnitude at dissociation compared to equilibrium. This observation is parallel to the idea of Boltzmann entropy  in the microcanonical ensemble in that the eneumeration of microstates (i.e.\ configurations) needed to represent the state increases the entropy. In this context, the multiconfigurational nature of the wavefunction is, to a degree, captured by this correlation proxy and introduces further uncertainty in the density matrix. In other words, this is a manifestation of the MEP. It can be posited that with Jaynes entropy maximization, under system constraints, we may converge towards the true ODM.\cite{Smith2002}

\begin{figure}
    \centering
    \includegraphics[width=\linewidth]{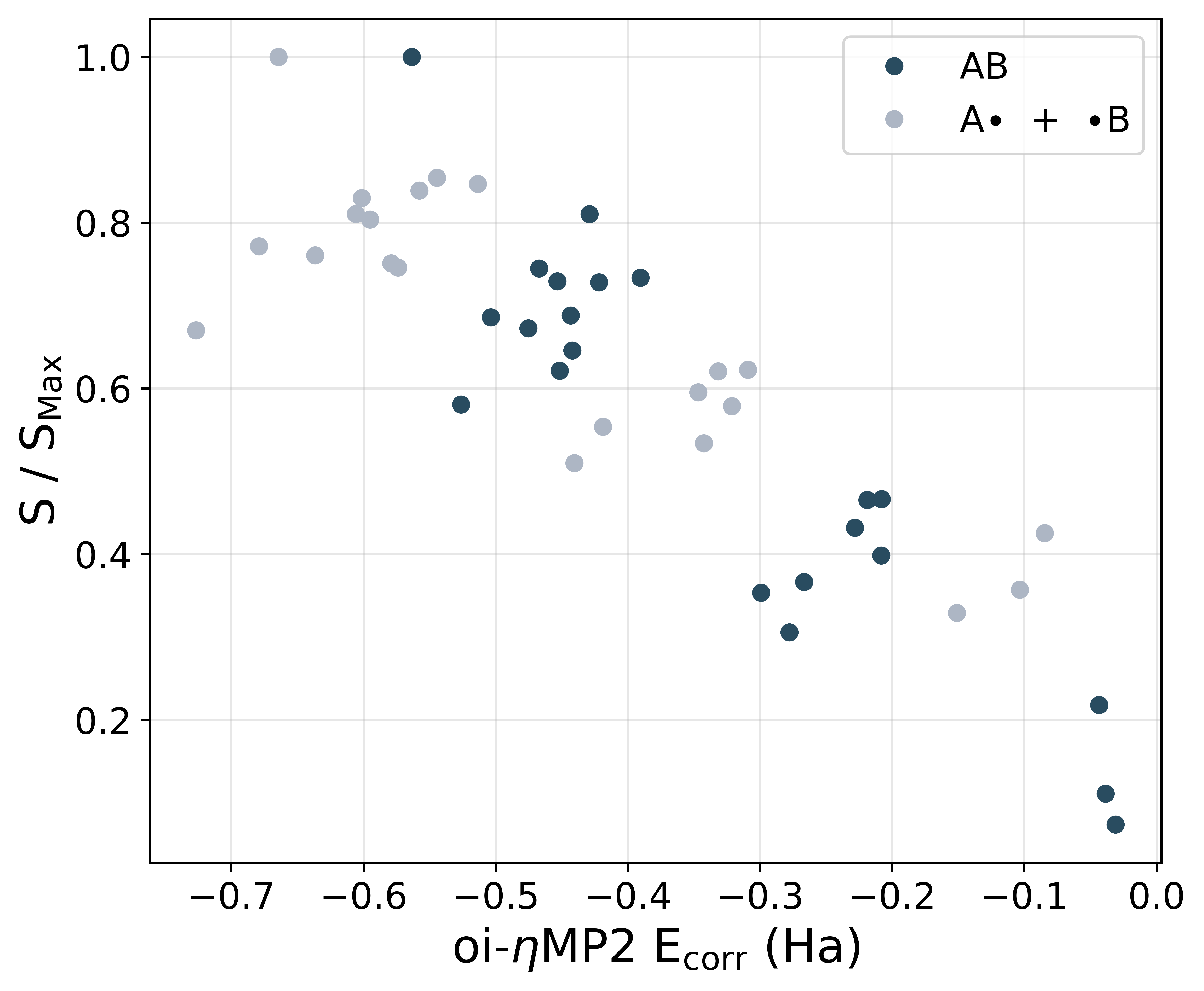}
    \caption{Comparison of second-order correlation energies $E_\mathrm{corr}$ from orbital-invariant (oi) $\eta \mathrm{MP2}$ and corresponding entropies normalized to the maximum value within each dataset for molecules at equilibrium and fragments separated at 100\r{A}. The parameter $\eta$ is tuned to provide  HOMO-LUMO occupations close to 1 for fragments stretched to 100\r{A}.}
    \label{fig:Ecorr-SJaynes-diradical-oi}
\end{figure}

In Fig.\ \ref{fig:Ecorr-SJaynes-diradical-oi} we inspect the relationship between $E_\mathrm{corr}$ from $\eta \mathrm{MP2}$ and dataset-normalized entropies at equilibrium and dissociation. Overall, the relationship between the entropy and $|E^{(2)}_{\mathrm{MP}}|$ is not observed to be strictly linear in a one-to-one sense. Though linearity may be achieved with $E_\mathrm{corr}$ renormalization and system-dependent fitting procedures, it is not needed to establish a useful heuristic.\cite{Ziesche1999,Lin2024}  On a per-system basis, $S_\mathrm{Jaynes}$ does appear to monotonically increase with the simultaneous additions of $E^{(2)}_{\mathrm{MP}}$ \textit{and} proper reconstruction of the $\eta \mathrm{MP2}$ ODM at molecular fragmentation. 

MP2 NOs of closed-shell, single-reference molecules can be used as initial guess orbitals for multiconfigurational SCF calculations while the magnitude of MP2 NOONs can be used to select approximate active spaces.\cite{Jensen1988a,Jensen1988b} In Table \ref{CAS-eta}, the quality of LiH NOONs from $\eta \mathrm{MP2}$ is examined in comparison to those of the complete active space SCF model (CASSCF) and active space configuration interaction (CASCI) over different orbital correlation windows along with their respective values for $S_\mathrm{Jaynes}$. CASSCF was performed with Q-Chem and ORCA\cite{Orca} was used to perform CASCI.\cite{Kollmar2019} 

While oi-$\eta$MP2 Jaynes entropies for LiH are overestimated with respect to CASSCF entropies, they approach those of full CI. Furthermore, the $\eta \mathrm{MP2}$ $n_\mathrm{HOMO}$ and $n_\mathrm{LUMO}$ at $\mathrm{r}_{100\text{\r{A}}}$ are close to the expected diradical-like occupations predicted by CASSCF and CASCI.
An even closer fit towards the ideal ONs could have been obtained with an input $\eta$ mantissa larger than the $10^{-4}$ Ha limit that was set. 
% Though $\eta \mathrm{MP2}$ does not use optimized orbitals, in contrast to CASSCF, its entropies at $\mathrm{r}_\mathrm{eq}$ and $\mathrm{r}_{100\text{\r{A}}}$ are qualitatively similar to CASSCF. The $\eta \mathrm{MP2}$ NOON representing the HOMO at $r_\mathrm{eq}$ is larger than the corresponding CASSCF NOON whereas the $\eta \mathrm{MP2}$ NOON representing the LUMO is smaller.

We also examined $S_\mathrm{Jaynes}$ and occupation structure for LiH using non-invariant $\eta$MP2. With oi-$\eta \mathrm{MP2}$, both occupied and virtual orbital energies are shifted in an even-handed way, while non-invariant $\eta$MP2 appears only to affect occupied energy levels. Consequently, we find that the NOONs obtained with oi-$\eta$MP2 are slightly over-stratified across the virtual space, leading to Jaynes entropies that are generally larger than those predicted from non-invariant $\eta$MP2. In either scheme, $\eta$MP2 offers reasonable total energies and practical reconstruction of the ODM.

%\section{Conclusions and Outlook}

By adopting Collins' conjecture as a \textit{general} rule, we have presented a method to address strong correlation in single-reference
perturbation theory. 
% Through re-expression of the second-order energy in terms of the Jaynes entropy and correlation parameter $\eta$, 
We demonstrate that $\eta \mathrm{MP2}$, which includes both static and dynamic correlation effects, can reconstruct the ODM to give the qualitatively correct electronic structure at the infinite separation limit by means of a ``density-tuning'' parameter, $\eta$ and an $\eta$-dependent multiplicative constant that scales the entropy-based static correlation contribution, $\gamma$. In stark contrast to RMP2, which diverges, contaminating the positive definiteness of the ODM, $\eta \mathrm{MP2}$ employs an entropy-driven renormalization of the one-particle orbital energies that corrects these deficiencies. In fact, BDEs obtained with orbital-invariant $\eta \mathrm{MP2}{-}\gamma S$ can approach the performance of GVB. In conjunction with MP2 or with other correlated wave function theories, a lever such as $\eta$ in $\eta$MP2 that is tuned to the ODM rather than to energies may be useful in capturing qualitatively correct electron distributions in molecules that exhibit strong correlation. Crucially, the associated electronic entropy of the ODM is obtained in an {\em ab initio} manner and can be used to efficiently account for static correlation. This essential step of building-in the static correlation component is missing from other regularized MP2 procedures, which merely discard poorly-described correlations.
% Different partitions of $\mathcal{H}$ that include balanced terms for virtual-orbital correlations may offer improvements to $\eta$-based methods.

The original proposition by Collins is again shown to be a practical conduit for incorporating of static correlation that can be supplemented by additional short-range electron-electron interaction terms via perturbation theory.
This work represents a foundational effort to formalize Collins' conjecture as an {\em ab initio} theory framed in terms of a first-order wavefunction whose corresponding ODM gives rise to non-integer occupation numbers, rather than obtaining them (and the associated entropy) from phenomenological smearing functions.
%Overall, our promising results with $\eta$MP2 suggest that additional research into the connection between Collins' conjecture and truly {\em ab initio} quantum chemistry methods is well warranted.

\section*{Supplementary Material}
See the Supplementary Material for details about geometries, basis sets, the non-invariant
approximation to $\eta$MP2, along with all occupation numbers, energies,
and entropies reported herein. Dipole calculations and occupation number scans for LiH using Fermi-Dirac smearing are also included
alongside additional BDE data for iron carbonyl species.

\begin{acknowledgments}
The authors would like to thank Sylvia J. Bintrim for her assistance with validating a nascent sandbox implementation of the method presented. This research was supported by the University of Pittsburgh and the University of Pittsburgh Center for Research Computing, RRID:SCR\_022735, through funding and computational resources.
Specifically, this work used the H2P cluster, which is supported by NSF award number OAC-2117681.
\end{acknowledgments}

\section*{Conflict of Interest}
The authors have no conflicts of interest to disclose.

\section*{Author Contributions}
\textbf{Abdulrahman Y. Zamani:} Data curation (equal); Formal analysis (equal); Software (equal); Writing -- original draft (lead); Writing -- review \& editing (equal).
\textbf{Kevin Carter-Fenk:} Data curation (equal); Formal analysis (equal); Software (equal); Conceptualization (lead); Supervision (lead); Writing -- review \& editing (equal).

\section*{Data Availability}
All data required to reproduce the results
within this article are included in the Supplementary Material.

% \appendix

% \section{Appendixes}

% test

% \section{A little more on appendixes}

% test

% \subsection{\label{app:subsec}A subsection in an appendix}

% test

% \nocite{*} %AZ
%\bibliography{apssamp}

\providecommand{\noopsort}[1]{}\providecommand{\singleletter}[1]{#1}%

\end{document}

%\preprint{APS/123-QED}

\title{\textbf{Supplementary Material for ``Towards \textit{ab initio} realizations of Collins' Conjecture''} 
}% 

\author{Abdulrahman Y. Zamani}
\author{Kevin Carter-Fenk}%
% \email{kay.carterfenk@pitt.edu}
\affiliation{%
Department of Chemistry, University of Pittsburgh,
Pittsburgh, Pennsylvania 15260, USA
}%

%  \homepage{http://www.Second.institution.edu/~Charlie.Author}
% \affiliation{
%  First affiliation for this author
% }%
% \affiliation{
%  second institution for this author
% \affiliation{%
%  Authors' institution and/or address\\
%  This line break forced with \textbackslash\textbackslash
% }%

% \collaboration{CLEO Collaboration}%\noaffiliation

\date{\today}% It is always \today, today,

\maketitle

%\tableofcontents

% \begin{figure}
%     \caption{Nice one caption!}\label{Fig:1}
%     Here is a nice figure
% \end{figure}

\SupplementaryMaterials
\section{Computational Details}
Bond dissociation energies (BDEs) or $D_e$ are inferred from experimental enthalpies $D_0$. A selection of molecules and reference data for these quantities were taken from earlier studies.\cite{Kraka1992a,Kraka1992b,Feng2003} Additional chemical species such as PH$_3$\cite{ph3}, H$_2$\cite{h2}, H$_2$S\cite{h2s}, and LiF\cite{lif} are included. Geometries computed at CCSD(T)/Aug-cc-pVTZ are obtained from the NIST Computational Chemistry Comparison and Benchmark Database\cite{Nist} (CCCBDB). Structures not available on NIST were computed at the same level of theory in a development version of Q-Chem.\cite{qchem} If  $D_e$ (non-measurable) reference values are not available, they are computed by removing zero-point energy (ZPE) differences between reactant and product from $D_0$.\cite{Cremer2000} ZPEs are computed with experimental frequencies obtained from NIST for reactants and products; if unavailable, frequencies are computed with CCSD/Aug-cc-pVTZ in Q-Chem.\cite{Besler1988} Theoretical BDEs calculated with unrestricted CCSD/cc-pVTZ and CCSD(T)/cc-pVTZ are obtained by taking the energy difference between the reactant species at equilibrium and the sum of fragmented products. The $\eta$-based methods use the cc-pVTZ basis set and BDEs are calculated from the  energy difference of the molecule at equilibrium and the atomic or relaxed/optimized molecular fragments separated at 100\r{A}. The restricted Hartree-Fock (HF) solutions for the dissociated species are tested for wavefunction stability. Post-HF calculations correlate all electrons. 

For obtaining proper diradical occupations, the input parameter $\eta$, with a mantissa of up to $10^{-4}$, is chosen such that the difference between the HOMO and LUMO $\eta \mathrm{MP2}$ NOONs at 100\r{A} is minimal.  The optimal $\eta$ for the fragmented molecule is then used on the same molecule at equilibrium. The value of $S_\mathrm{Jaynes}$ is computed with the sum of up-spin $\alpha$ and down-spin $\beta$ entropies and applied to $\mathrm{HF}{-}\eta S$ and $\eta \mathrm{MP2}{-}\eta S $ along with the same individually optimized entries for $\eta$. The optimal value of $\gamma$ is the determined $\eta$ scaled to obtain the lowest average of errors against experiment and theory.
\section{L\MakeLowercase{i}H Dissociation with Fermi-Dirac Occupations}
\newpage
Fermi-Dirac (F-D) occupations of LiH using the original (old) and reformulated (r) TAO methods\cite{Chai2012,Yeh2022} with various reference determinants and the cc-pVTZ basis set. Convergence of stable SCF solutions with LDA at $r_{100\text{\r{A}}}$ was difficult, so $r_{10\text{\r{A}}}$ was used instead.

\begin{figure}[H]
\centering
  \label{Fig:lih} 
    \includegraphics[width=\linewidth]
    {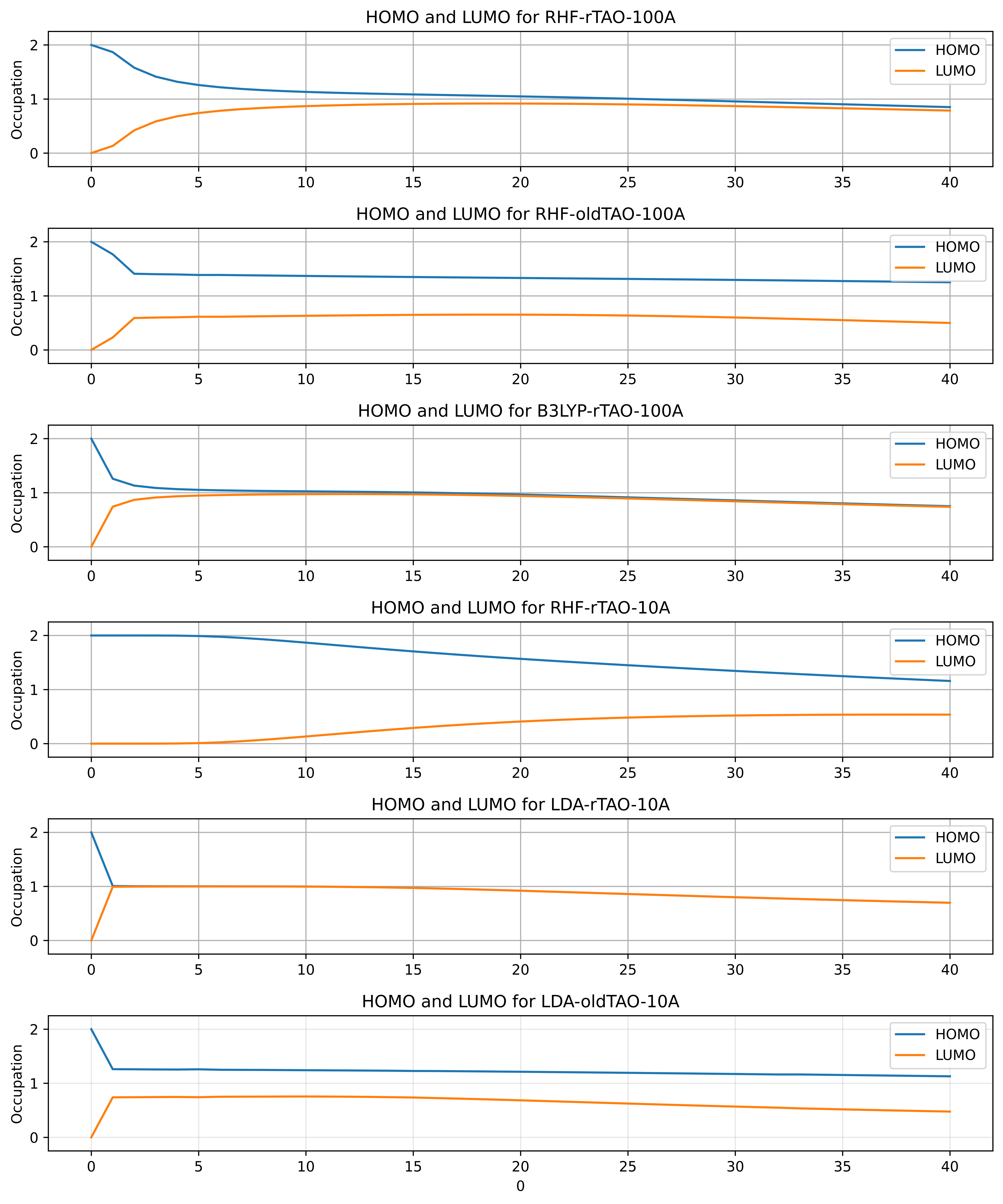}
\end{figure}

% \section{Parametrization of $\eta$ to ``Ionic'' Occupations}
\section{Results for Parametrization of $\eta$ with Non-Orbital-Invariant $\eta$MP2}

Obviously many possible constructions of non-orbital-invariant construction of $\eta$MP2 exist.
The particular choice that we made involves taking $\mathcal{H}_0$ as introduced in Eq.~17 of the main text
to be diagonal:
\begin{align}
    \mathcal{H}_0 &= \sum_p \bigg(\mathcal{F}_{pp}  a_{p}^{\dagger} a_{p} - \eta  s_p\bigg) \nonumber \\
    &\approx \sum_{ij}\bigg(f_i^j+\eta v_{ab}^{kj}t_{ik}^{ab}\bigg)a_j^\dagger a_i+\sum_{ab}\bigg(f_b^a-\eta v_{cb}^{ij}t_{ij}^{ac}\bigg)a_b^\dagger a_a \nonumber \\
    &\approx \sum_{ij}\bigg(f_i^j+\eta v_{ab}^{kj}t_{ik}^{ab}\bigg)\delta_{ij}+\sum_{ab}\bigg(f_b^a-\eta v_{cb}^{ij}t_{ij}^{ac}\bigg)\delta_{ab} \nonumber \\
    &= \sum_{i}F_{ii}+\sum_{a} F_{aa}
\end{align}
Clearly, this approximation that takes only diagonal elements to be non-zero in the canonical orbital basis
leads immediately at the typical expression for the MP2 amplitudes,
\begin{align}
    \Big[\big(F_c^a\delta_d^b+\delta_c^aF_d^b\big)\delta_i^k\delta_j^l-\big(F_i^k\delta_j^l+\delta_i^kF_j^l\big)\delta_c^a\delta_d^b\Big]t_{kl}^{cd} &= -v_{ij}^{ab}\\
    (\tilde{\varepsilon}_a+\tilde{\varepsilon}_b-\tilde{\varepsilon}_i-\tilde{\varepsilon}_j)t_{ij}^{ab} = -v_{ij}^{ab}\\
    t_{ij}^{ab} = -\frac{v_{ij}^{ab}}{\tilde{\varepsilon}_a+\tilde{\varepsilon}_b-\tilde{\varepsilon}_i-\tilde{\varepsilon}_j}
\end{align}
The mosaic terms (the threefold contractions over integrals and amplitudes added to the one-particle Fock matrix) may be interpreted as the correlation component of pair removal energies and pair relaxation energies respectively.\cite{Pickup1973,Szabo} Thus, the orbital energies in this non-invariant partitioning (and the reason for its non-invariance) are shifted by an incomplete notion of the pair-correlation energies as not all of the information stored in the amplitudes is correctly used. This diagonal approximation therefore destroys information, imparting the undesirable property of dependence on the particular choice of orbital basis. Hence, we strongly recommend against such a naive shifting of orbital energies even though we explore the results given by such an approach in the main text.

We instead adopt the philosophy that the underlying electronic structure method responsible for capturing dynamical correlation
should be orbital invariant and information conserving. This way, the resultant $E_{\text{corr}}=E_{\text{dyn}}+E_{\text{static}} = E_{\text{dyn}} - \gamma S$
approach is invariant up until the entropy contribution is added in the natural orbital basis.

\begin{figure}[H]
    \centering
    \includegraphics[width=0.8\linewidth]{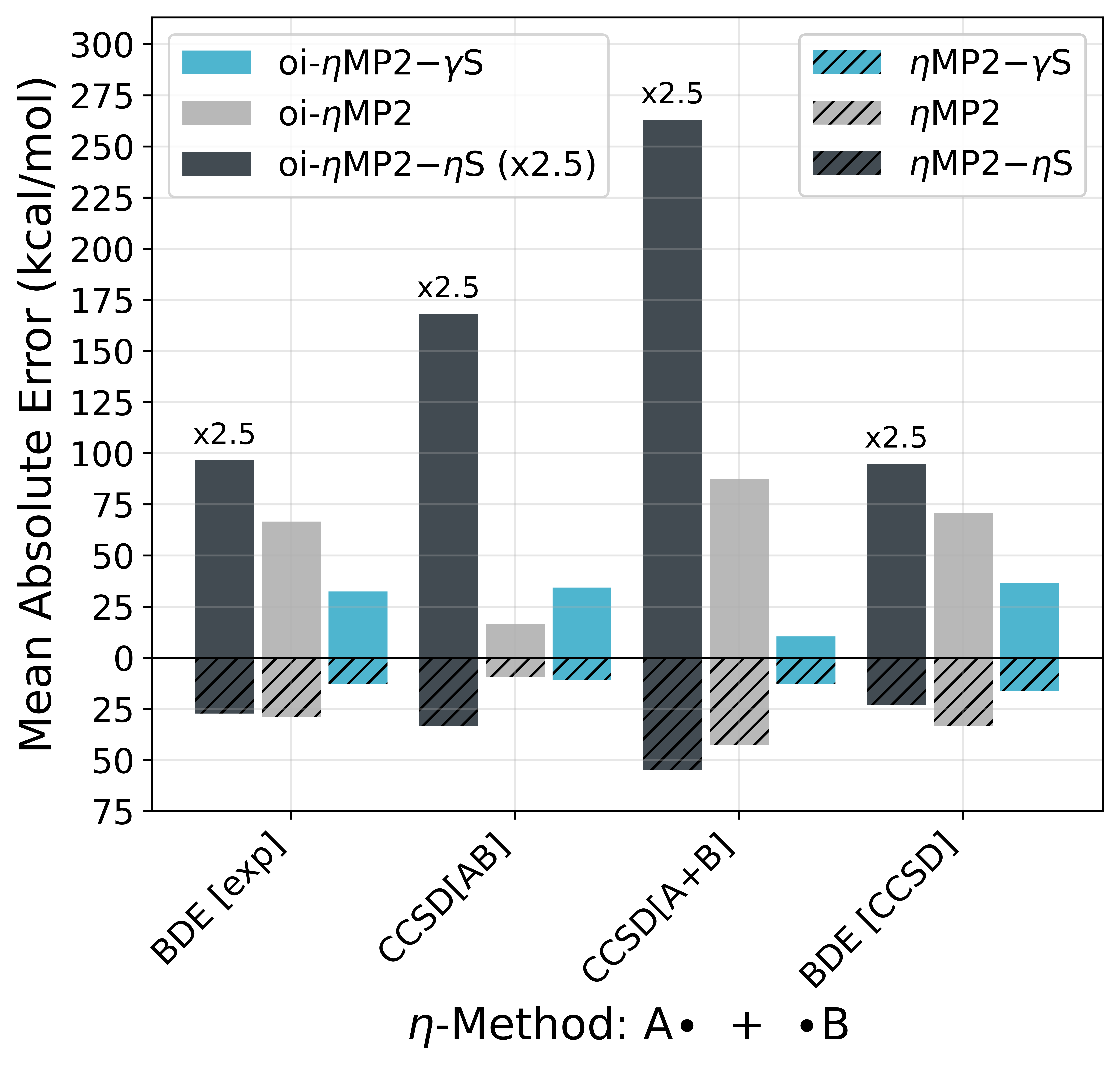}
    \caption{Average errors in kcal/mol of orbital-invariant (oi) and non-invariant $\eta$-based methods with respect to CCSD total energies, CCSD BDEs, and experimental BDEs. The parameter $\eta$ is tuned to provide  HOMO-LUMO occupations close to 1 for fragments stretched to 100\r{A}. The optimal value of oi and non-oi $\gamma$ is $\eta$ scaled by 0.116 and 0.476 respectively.}
    \label{fig:shared-eta-diradical-ccsd}
\end{figure}
\begin{figure}[H]
    \centering
    \includegraphics[width=0.8\linewidth]{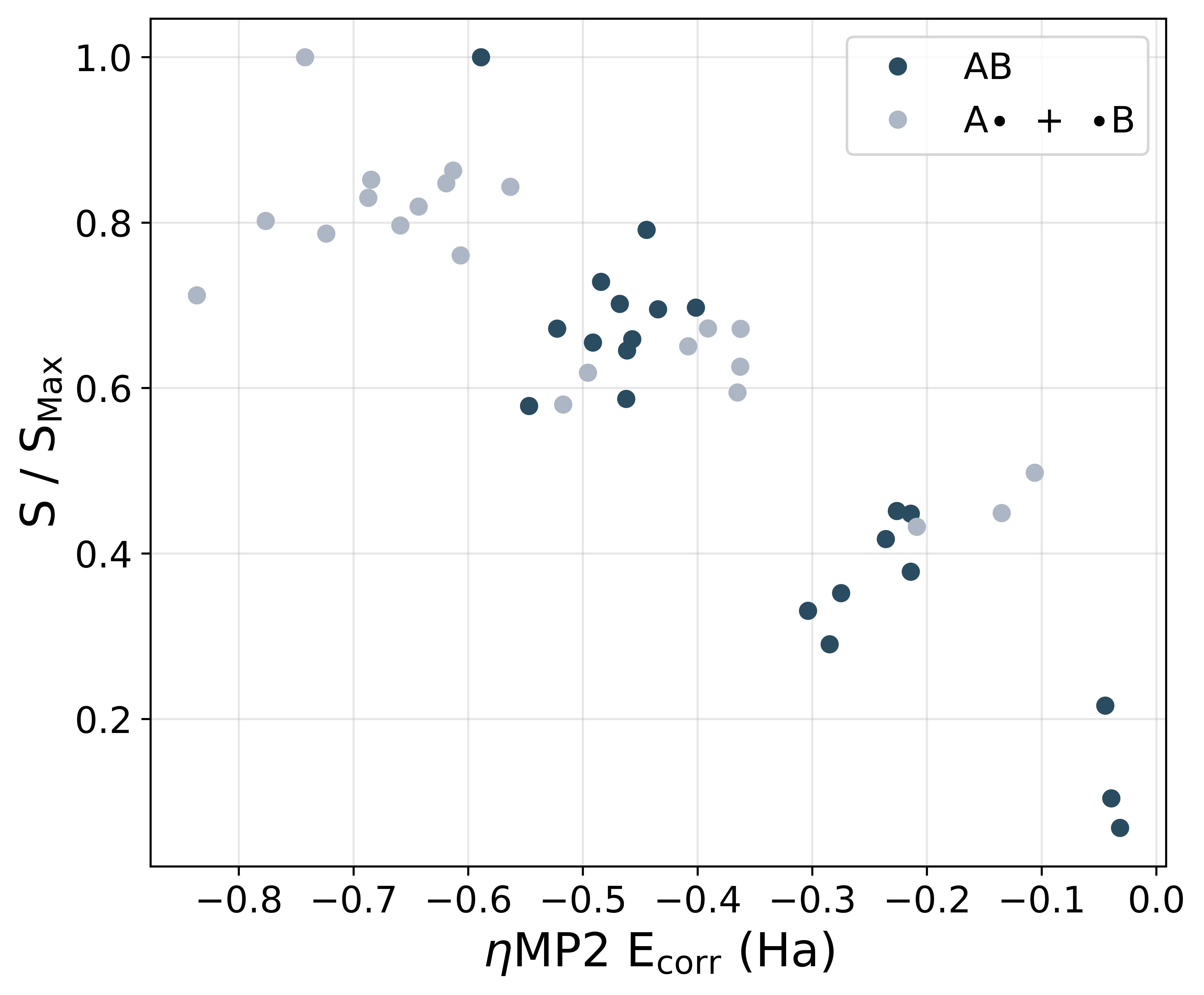}
    \caption{Comparison of second-order correlation energies $E_\mathrm{corr}$ from $\eta \mathrm{MP2}$ and corresponding entropies normalized to the maximum value within each dataset for molecules at equilibrium and fragments separated at 100\r{A}. The parameter $\eta$ is tuned to provide  HOMO-LUMO occupations close to 1 for fragments stretched to 100\r{A}.}
    \label{fig:Ecorr-SJaynes-diradical}
\end{figure}
\newpage
\begin{table}[ht]
\centering
\caption{$S_\mathrm{Jaynes}$ and NOONs from RI-$\eta \mathrm{MP2}$/cc-pVTZ and CASCI/cc-pVTZ with growing active spaces $(N,M)$ for LiH calculated at equilibrium ($\mathrm{r}_\mathrm{eq}$)  and stretched ($\mathrm{r}_{100\text{\r{A}}}$) bond lengths. CASSCF/cc-pVTZ results are in \textit{italics}.}\vspace{3pt}
\setlength{\tabcolsep}{5pt}  %5pt
\renewcommand{\arraystretch}{1.00} %1.25
\begin{tabular}{l|ccccccc}
\hline\hline
$\mathrm{r}_\mathrm{eq}$  & $\eta \mathrm{MP2}$ & (4,4) & (4,8) & (4,16) & (4,32) & (4,44) \\ 
$S_\mathrm{Jaynes}$ & 0.224 & 0.002 & 0.025 & 0.155 & 0.425 & 0.467  \\
&  & \textit{0.252} & \textit{0.337} & \textit{0.358} & \textit{0.357} & \textit{0.354}  \\
 $n_\mathrm{HOMO}$ & 1.971 &  2.000 & 1.997 &  1.980 & 1.942 & 1.941 \\ 
 &  & \textit{1.952} & \textit{1.941} & \textit{1.941} & \textit{1.942} & \textit{1.942}  \\
 $n_\mathrm{LUMO}$ & 0.012 &  0.000 & 0.000 &  0.003 & 0.005 & 0.004 \\
 &  & \textit{0.037} & \textit{0.031} &  \textit{0.030} & \textit{0.030} & \textit{0.029} 
 \\ \cline{1-7} 
 $\mathrm{r}_{100\text{\r{A}}}$  & $\eta \mathrm{MP2}$ & (4,4) & (4,8) & (4,16) & (4,32) & (4,44) \\ 
 $S_\mathrm{Jaynes}$ & 1.528 &  1.386 & 1.386 & 1.389 & 1.631 & 1.673 \\
 &  & \textit{1.409} & \textit{1.409} & \textit{1.409} & \textit{1.409} & \textit{1.409}  \\
 $n_\mathrm{HOMO}$ & 0.991 &  1.000 &  1.000 & 1.000 & 0.935 &  0.930\\ 
 &  & \textit{1.000} &  \textit{1.000} &  \textit{1.000} &  \textit{1.000} &  \textit{1.000}  \\
 $n_\mathrm{LUMO}$ & 0.992 & 1.000 & 1.000 &  1.000 &  1.000 &  1.000 \\ 
  &  & \textit{1.000} &  \textit{1.000} &  \textit{1.000} &  \textit{1.000} &  \textit{1.000} \\
\hline\hline
\end{tabular}
\label{CAS-eta}
\end{table}
%

%Fe(CO)5 
\begin{table}[ht]
\centering
\caption{Computed BDEs (kcal/mol) for singlet dissociation pathways of  iron carbonyl clusters. The correlation entropy parameters for the orbital invariant (oi) $\eta$MP2 methods use $\eta$ = 384 mHa and $\gamma$ = 45 mHa. Restricted (R) MP2, $\eta$MP2, and CCSD calculations use the RI approximation and cc-pVTZ basis. Structures optimized with $\omega$B97M-V/pcseg-1 are taken from ref. \cite{ironComplex} or computed with the same model chemistry. Signed errors relative to CCSD results are in parenthesis.}\vspace{3pt}
\setlength{\tabcolsep}{5pt}  %5pt
\renewcommand{\arraystretch}{1.00} %1.25
\begin{tabular}{l|cccc}
\hline\hline
Dissociation Path & RMP2 & oi-$\eta \mathrm{MP2}$ & oi-$\eta \mathrm{MP2}-\gamma S$ \\ 
$\mathrm{Fe(CO)_5} \rightarrow \mathrm{Fe(CO)_4}  [\mathrm{C_{2v}}] + \mathrm{CO}$ & 68.2 (14.3) & 57.4 (3.6) & 63.6 (9.7)  \\
$\mathrm{Fe(CO)_5} \rightarrow \mathrm{Fe(CO)_4}  [\mathrm{C_{3v}}] + \mathrm{CO}$ & 89.8 (36.4) & 70.8 (17.4) & 82.0 (28.7) \\
 $\mathrm{Fe(CO)_5} \rightarrow \mathrm{Fe(CO)_4}  [(\mathrm{C_{3v}+C_{2v}})/2]^{a} + \mathrm{CO}$ & 79.0 (25.4) & 64.1 (10.5) & 72.8 (19.2)  \\ 
 $\mathrm{Fe(CO)_4} [\mathrm{C_{2v}}] \rightarrow \mathrm{Fe(CO)_3}  [\mathrm{C_{s}}] + \mathrm{CO}$ & 82.7 (33.5) & 66.2 (17.1) & 76.4 (27.2) \\
  $\mathrm{Fe(CO)_4} [\mathrm{C_{3v}}] \rightarrow \mathrm{Fe(CO)_3}  [\mathrm{C_{s}}] + \mathrm{CO}$ & 61.1 (11.5) & 52.9 (3.2) & 58.0 (8.3) \\
   $\mathrm{Fe(CO)_4}  [(\mathrm{C_{3v}+C_{2v}})/2]^{a} \rightarrow \mathrm{Fe(CO)_3}  [\mathrm{C_{s}}] + \mathrm{CO}$ & 71.9 (22.5) & 59.6 (10.2) & 67.2 (17.8) \\ 
\hline\hline 
\end{tabular}
\label{feco5}
\\ \flushleft{$^{a}$The $\mathrm{C_{2v}}$ and $\mathrm{C_{3v}}$ $\mathrm{Fe(CO)_{4}}$ structures have a small barrier for interconversion through a scissoring motion about $\mathrm{CO}$ ligand pairs. The average of energetic and entropic quantities of stretched $\mathrm{Fe(CO)_4} + \mathrm{CO}$ and equilibrium $\mathrm{Fe(CO)_4}$ for both symmetries are used to compute average BDEs. 
}
\end{table}
\newpage
\section*{Equilibrium Electric Properties: Dipole Moments}
%dipole moments
\begin{table}[ht]
\centering
\caption{Computed dipole moments using the cc-pVTZ basis set alongside the RI approximation. All dipoles are expressed in units of Debye.
Absolute percent error is displayed in parenthesis.}\vspace{3pt}
\setlength{\tabcolsep}{5pt}  %5pt
\renewcommand{\arraystretch}{1.00} %1.25
\begin{tabular}{lc|ccc}
\hline\hline
System & Experiment & oi-$\eta \mathrm{MP2}-\gamma S^b$ & MP2 & CCSD \\ 
CO & 0.12 & 0.12 (0.0) & 0.32 (161.0) & 0.11 (10.82)\\
O$_3$ & 0.53 & 0.68 (28.3) & 0.45 (15.7) & 0.59 (11.9)\\
HBS & 1.30 & 1.42 (9.4) & 1.28 (1.3) & 1.40 (8.2)\\
HCNO & 3.06 & 2.96 (3.3) & 2.48 (19.0) & 2.97 (2.8)\\
NaLi & 0.46 & 0.57 (23.9) & 0.55 (19.0) & 0.75 (63.8)\\
BH$_2$Cl & 0.75 & 0.61 (18.7) & 0.59 (21.4) & 0.61 (18.3)\\
SF$_2$ & 1.05 & 1.24 (18.1) & 1.14 (8.3) & 1.21 (15.3)\\
SiO & 3.10 & 3.05 (1.6) & 2.70 (12.7) & 3.03 (2.3)\\
SO$_2$ & 1.63 & 1.80 (10.4) & 1.58 (3.1) & 1.77 (8.5)\\
HCCF & 0.73 & 0.65 (11.0) & 0.60 (17.3) & 0.69 (4.9)\\
ClF & 0.88 & 1.07 (21.6) & 0.95 (8.5) & 1.01 (14.9)\\
BH$_2$F & 0.82 & 0.77 (6.1) & 0.72 (12.0) & 0.77 (5.9)\\
BHF$_2$ & 0.97 & 0.91 (6.3) & 0.86 (11.3) & 0.91 (5.9)\\
HF & 1.83 & 1.85 (1.3) & 1.83 (0.0) & 1.83 (0.0)\\
\hline
\multicolumn{2}{c}{MAPE} & 11.4 & 22.2 & 12.4\\
\multicolumn{2}{c}{Max \% Error} & 28.3 & 161.0 & 63.8\\
\hline\hline 
\end{tabular}
\label{dipoles}
\\ \flushleft{$^{a}$Experimental values and geometries obtained from Ref.~\cite{HaiHea18} except for HF which was obtained from NIST.}
\\ \flushleft{$^{b}$Using $\eta=1.725$ optimized to best reproduce the dipole moment of CO. A choice of $\alpha=1/(22.4\eta)$ maintains the same ratio between $\eta$ and $\gamma$ reported for BDEs in the main text.
}
\end{table}

%   A reference to the nice figure: see \ref{Fig:1}

%   A reference to the nicer figure: see \ref{Fig:2}

% \appendix

% \section{Appendixes}

% test

% \section{A little more on appendixes}

% test

% \subsection{\label{app:subsec}A subsection in an appendix}

% test

% \nocite{*} %AZ
\clearpage
\pagebreak
\bibliography{apssamp}% Produces the bibliography via BibTeX.